# Transfer Learning for Automated OCTA Detection of Diabetic Retinopathy

David Le[1], Minhaj Alam[1], Cham Yao[4], Jennifer I. Lim[2], R.V.P. Chan[2], Devrim Toslak[1,3], and Xincheng Yao[1,2,*]

[1]Department of Bioengineering, University of Illinois at Chicago, Chicago, IL, USA

[2]Department of Ophthalmology and Visual Sciences, University of Illinois at Chicago, Chicago, IL, USA

[3]Department of Ophthalmology, Antalya Training and Research Hospital, Antalya, Turkey

[4]Hindsdale Central High School, Hinsdale, IL, USA

*xcy@uic.edu

**Grant information:** National Institute of Health (NIH) (R01EY029673, R01EY030101, R01EY023522, and P30 EY001792); unrestricted grant from Research to Prevent Blindness; and Richard and Loan Hill endowment.

**Disclosure:** none.

**ABSTRACT**

**Purpose:** To test the feasibility of using deep learning for optical coherence tomography angiography (OCTA) detection of diabetic retinopathy (DR).

**Methods:** A deep learning convolutional neural network (CNN) architecture VGG16 was employed for this study. A transfer learning process was implemented to re-train the CNN for robust OCTA classification. In order to demonstrate the feasibility of using this method for artificial intelligence (AI) screening of DR in clinical environments, the re-trained CNN was incorporated into a custom developed GUI platform which can be readily operated by ophthalmic personnel.

**Results:** With last nine layers re-trained, CNN architecture achieved the best performance for automated OCTA classification. The overall accuracy of the re-trained classifier for differentiating healthy, NoDR, and NPDR was 87.27%, with 83.76% sensitivity and 90.82% specificity. The AUC metrics for binary classification of healthy, NoDR and DR were 0.97, 0.98 and 0.97, respectively. The GUI platform enabled easy validation of the method for AI screening of DR in a clinical environment.

**Conclusion:** With a transfer leaning process to adopt the early layers for simple feature analysis and to re-train the upper layers for fine feature analysis, the CNN architecture VGG16 can be used for robust OCTA classification of healthy, NoDR, and NPDR eyes.

**Translational Relevance:** OCTA can capture microvascular changes in early DR. A transfer learning process enables robust implementation of convolutional neural network (CNN) for automated OCTA classification of DR.

**INTRODUCTION**

As the leading cause of preventable blindness in working-age adults, diabetic retinopathy (DR) affects 40-45% of diabetic patients [1]. In United States (US) alone, the DR patients are estimated to increase from 7.7 million in 2010 to 14.6 million by 2050 [1]. Early detection, prompt intervention, and reliable assessment of treatment outcomes are essential to prevent irreversible visual loss from DR. With early detection and adequate treatment, more than 95% of DR related vision loss can be preventable [2]. Retinal vascular abnormalities, such as microaneurysms, hard exudates, retinal edema, venous beading, intraretinal microvascular anomalies and retinal hemorrhages are common DR findings [3]. Therefore, imaging examination of retinal vasculature is important for DR diagnosis and treatment evaluation. Traditional fundus photography provides limited sensitivity to reveal subtle abnormality correlated with early DR [4-7]. Fluorescein angiography (FA) can be used to improve imaging sensitivity of retinal vascular distortions in DR [8,9], but FA requires intravenous dye injections which may produce side effects and requires following monitoring and management carefully. Optical coherence tomography angiography (OCTA) provides a noninvasive method for better visualization of retinal vasculatures [10]. OCTA allows visualization of multiple retinal layers with high resolution, and thus it is more sensitive than FA in detecting subtle vascular distortions correlated with early eye conditions [11,12].

Recent development of quantitative OCTA opens a unique opportunity to enable computer-aided disease detection and AI classification of eye conditions. Quantitative OCTA analysis has been explored for objective assessment of DR [13-16], age-related macular degeneration (AMD)[17,18], vein occlusion (VO) [19-22], SCR [23-25], etc. Supervised machine learning has also recently validated for multiple-task classification to differentiate control, DR and SCR from each other [26]. In principle, unsupervised deep learning may provide a simple solution to foster clinical deployment of AI classification of OCTA images. In ophthalmology, deep learning has been applied for segmentation of the retinal layers [27] and avascular area [28] in OCT. The common deep learning algorithms used for this AI renaissance are convolutional neural networks (CNNs). Inspired by the human brain and visual pathway, CNNs contain millions of artificial neurons (also referred to as parameters) to process image features in a feed-forward process, i.e., extracting and processing simple features in early (bottom) layers and complex features in later (upper) layers [29]. To train a CNN for a specific classification task requires millions of images to optimize the network parameters [30]. For everyday tasks, such as classifying cats and dogs, there are abundant available images to meet the data requirement for the training of the CNN classifier. However, for new imaging modality, such as the OCTA, the limitation of available image poses as an obstacle for practical implementation of deep learning.

In order to overcome the limitation of data size, a transfer learning approach has been demonstrated for deep learning. Transfer learning is a training method to adopt some weights of a pre-trained CNN and appropriately re-train the CNN to optimize the weights for a specific task, i.e., AI classification of retinal images [31]. In fundus photography, transfer learning has been explored to conduct artery-vein segmentation [32], glaucoma detection [33,34], and diabetic macular thinning assessment [35]. Recently, transfer learning has also been explored in OCT for detecting choroidal neovascularization (CNV) and diabetic macular edema (DME) [31], and AMD [36].

In principle, transfer learning can involve a single layer or multiple layers, because each layer has weights that can be re-trained. For example, the specific number of layers required for re-training in a 16-layer CNN (Fig. 1) may vary, depending on the available dataset and specific task interested. Moreover, compared to traditional fundus photography and OCT, deep learning in OCTA classification is still unexplored due to the limited size of publicly available datasets. In this study, we propose to employ transfer learning to train a CNN classifier for DR detection in OCTA, and to provide a custom graphical user interface (GUI) platform to verify the potential of clinical deployment of the AI classification of DR using OCTA.

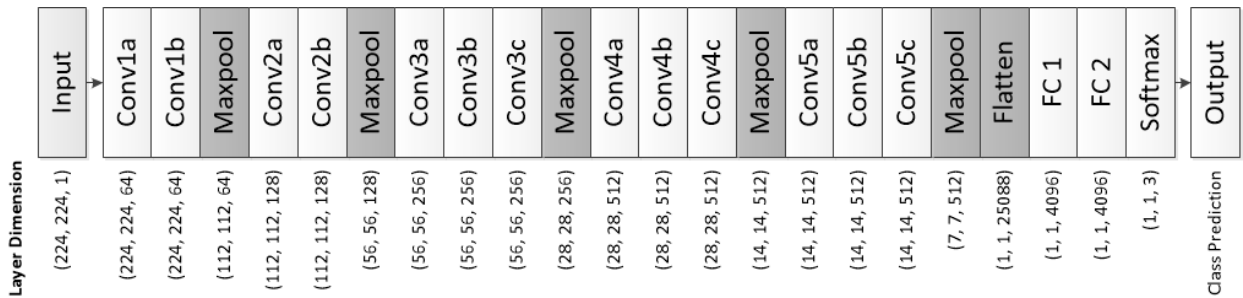

*Figure 1 The deep learning CNN used for OCTA DR detection is VGG16, a network that contains 16 trainable layers (convolution (Conv) and fully connected layers (FC)). Each layer has its corresponding output layer dimensions below each block. All convolution and fully connected (FC) layers are followed by a ReLU activation function. The softmax*

*layer is a fully connected layer that is followed by a softmax activation function. Maxpool and Flatten layers are operational layers with no tunable parameters.*

## METHODS

This study is in adherence to the ethical standards present in the Declaration of Helsinki and was approved by the institutional review board of the University of Illinois at Chicago (UIC).

**Data Acquisition**

6 x 6 mm$^2$ FOV OCTA data were acquired using an ANGIOVUE spectral domain (SD) OCTA system (Optovue, Fremont, CA) with a 70-kHz A-scan rate, a lateral resolution of ~15 um, and an axial resolution of ~5 um. All OCTA images were qualitatively examined for severe motion or shadow artifacts. Images with significant artifacts were excluded for this study. OCTA data was exported using ReVue (Optovue) software and custom-developed Python procedures were used for image processing.

**Patient Demographics**

Subjects and DM patients with and without DR were recruited from the UIC retina clinic. The patients present in this study are representative of a university population of DM patients who require clinical diagnosis and management of DR. Two board-certified retina specialists classified the patients based on the severity of DR according to the Early Treatment Diabetic Retinopathy Study (ETDRS) staging system. All patients underwent complete anterior and dilated posterior segment examination (JIL, RVPC). All control OCTA images were obtained from healthy volunteers that provided informed consent for OCT/OCTA imaging. All subjects underwent OCT and OCTA imaging of both eyes (OD and OS). The images used in this study did not include eyes with other ocular diseases or any other pathological features in their retina such as epiretinal membranes and macular edema. Additional exclusion criteria included eyes with prior history of intravitreal injections, vitreoretinal surgery or significant (greater than a typical blot hemorrhage) macular hemorrhages. Subject and patient characteristics including sex, age, duration of diabetes, diabetes type, HBA1C, and hypertension prevalence are summarized in Table 1.

Table 1 Demographics of OCTA data

|  | **CONTROL** | **NODR** | **DR** |
|---|---|---|---|
| **Number Of Subjects** | 20 | 17 | 60 |
| **Sex (Male/Female)** | 12/8 | 6/11 | 34/26 |
| **Age (Mean +/- Sd), Years** | 42 +/- 9.8 | 66.4 +/- 10.14 | 52.91 +/- 10.45 |
| **Age Range, Years** | 25 - 71 | 49 - 86 | 24 – 74 |
| **Duration Of Diabetes (Mean +/- Sd), Years** | - | - | 19.71 +/- 11.93 |
| **Diabetes Type (% Type Ii)** | - | 100 | 100 |
| **Insulin Dependent (Y/N)** | - | 14/3 | 34/60 |
| **Hba1c, %** | - | 5.9 +/- 0.7 | 7.2 +/- 0.9 |
| **HTN Prevalence, %** | 10 | 17 | 41 |

**Classification Model Implementation**

The CNN architecture chosen in this study is VGG16 [37]. The network specifications and design are illustrated in Fig. 1. The pre-trained weights were from the ImageNet dataset [38]. Training was performed with a batch size of 12 images with stochastic gradient descent optimization using a learning rate of 0.0001. Each classifier was trained with early stopping, and experimentally the model converges within ~70 epochs, where model refers to the re-trained classifier. During each iteration, data augmentation in the form of random rotation, horizontal and vertical flips, and zoom were performed to prevent the model from overfitting. In addition, all images were resized from (304 x 304) pixel$^2$ to (224 x 224) pixel$^2$ to meet the network input specifications of VGG16.

**Statistics**

In this study, manual classifications were used as reference standard for determining the receiver operating characteristic (ROC) curves and area under the curve (AUC). AUC was used as an index of the performance of the model along with sensitivity (SE), specificity (SP), diagnostic accuracy (ACC), and misclassification error which can be determined as follows.

$$SE = TP/TP + FN \qquad (1)$$

$$SP = TN/TN + FP \qquad (2)$$

$$ACC = TP + TN / TP + TN + FP + FN \qquad (3)$$

$$\text{Misclassification Error} = FP + FN/TP + TN + FP + FN \qquad (4)$$

Where TP are true positives, the model correctly predicts a positive class, TN are true negatives, where the model correctly predicts a negative class. FP are false positives, where the model incorrectly predicts a negative class as positive, and FN are false negatives where the model incorrectly predicts a positive class as negative. To evaluate the performance of each model, fivefold cross-validation was implemented. The training dataset was randomly divided into five subsets, four subsets were used for training and one subset was used for validation.

**Transfer Learning and Model Selection**

In practice, it takes hundreds of thousands of data samples i.e. images to optimize the millions of parameters of a CNN. Training CNNs on smaller datasets will often lead to overfitting, where the CNN has memorized the dataset, i.e. has high performance when predicting on the same dataset and fails to perform well on new data [39]. Transfer learning, which leverages the weights in a pre-trained network, has been established to overcome the overfitting problem. Transfer learning is well suited for CNNs because CNNs extracts features in a bottom-up hierarchical structure. This bottom-up process is analogous to the human visual pathway system [40]. Where in the early layers of the CNN extracts simple features, such as lines and color; and in the later layers the CNN extracts complex features. Therefore, only the weights in later layers are generally required to be adjusted to achieve optimal performance. Furthermore, the more similar or related the target dataset to the pre-trained dataset, the fewer layers of the CNN are required for re-training. Since in this study, the pre-trained weights are optimized by the ImageNet dataset [38] and the target dataset are OCTA images have high dissimilarity, we conduct a transfer learning process to determine the appropriate number of re-trained layers required in a CNN to achieve robust performance of OCTA classification. For this study, misclassification error will be used for quantitative assessment of the CNN performance, and one standard deviation rule was used for model selection. The one standard deviation rule states that a model selected is within one positive standard deviation from the misclassification error of the best performing model [41]. The model, which requires the least number of re-trained layers for a comparable performance to the best trained model, will be selected.

**GUI/Software Integration**

The CNN classifier was trained, and the models were evaluated on Python 3.7.1 using the programming libraries Keras 2.24 and Tensorflow 1.13.1. Training was performed on a Windows 10 computer using a NVIDIA GeForce RTX 2080 Ti Graphics Processing Unit (GPU) with NVIDIA cuda (v10.0.0) and cudnn (v7.3.1) libraries (http://www.nvidia.com).The GUI was developed using Java 11.01. and Python 3.7.1 on a macOS computer using an Intel UHD Graphics 617 GPU. After evaluating the best performing model, the weights of the model were exported into an ~500mb file (.hdf5) and integrated with a java/python-based graphical user interface (GUI). Statistics were performed using Microsoft Excel (Microsoft Corporation, Redmond, WA).

**RESULTS**

This study included 24 eyes from 17 NoDR patients, 75 eyes from 60 DR patients, and 32 eyes from 20 control subjects. In order to minimize the performance variance from a limited dataset, an augmented

OCTA dataset was generated by applying random flips, rotations and zooms on our 131 OCTA images to generate 3930 OCTA images. In following discussion, the raw OCTA images (n = 131) will be referred as Dataset 1 and the augmented OCTA images (n = 3930) will be referred as Dataset 2.

**Model Selection**

The model selection process was used to identify the required number of re-trained layers for the best performance of OCTA classification. The layers in the VGG16 CNN were sequentially re-trained starting from the last to the beginning layers. Quantitative comparison between the misclassification errors for individual models, i.e., variable number of re-trained layers, is shown in Fig. 2. In this study, we evaluated each model with Dataset 1 and Dataset 2. This enables us to access the real-world performance with the raw OCTA images, and to verify the potential performance of each model on a larger dataset. Using the one standard deviation rule, which recommends choosing a less complex model that is within one standard deviation of the misclassification error of the best performing model. The model re-trained with nine layers meets the one standard deviation criteria for both dataset validations and was selected for following cross-validation study.

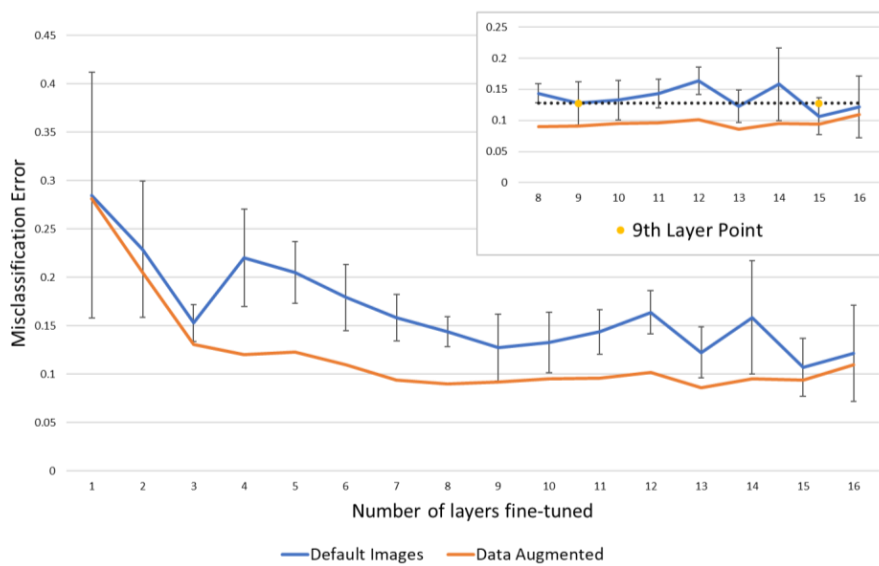

*Figure 2 A transfer learning performance study was conducted to determine how many layers is necessary for effective transfer learning in OCTA images. In our study, since Maxpooling, and Flattening operations have no optimizable parameters, only Convolutional and Dense layers will be numerically counted from last to first, i.e. one to sixteen. Therefore, in the transfer learning study, models re-trained with one to sixteen layers were compared. The blue line represents validation on raw OCTA images, and the orange line represents validation on augmented OCTA images. In the top-right hand corner, we zoomed in to compare the performance of the 9th and 15th layer. The yellow point is the misclassification error of the model re-trained with nine layers and the black dashed line illustrates that overall that the error of the re-trained nine-layer is within one positive standard deviation of the re-trained fifteen-layer model (the best performing model).*

**Cross-validation Study**

Using a fivefold cross-validation method, we evaluated the performance of the selected nine layers re-trained model with Dataset 1 and Dataset 2.

*Table 2 Cross Validation Multi-Label Confusion Matrix for Dataset 1*

| n = 131 | | **PREDICTED LABEL** | | |
|---|---|---|---|---|
| | | Control | NoDR | DR |
| **True Label** | Control | 25 | 3 | 4 |
| | NoDR | 1 | 23 | 0 |
| | DR | 9 | 8 | 58 |

*Table 3 Cross Validation Multi-Label Confusion Matrix for Dataset 2*

| n = 3930 | | **Predicted Label** | | |
|---|---|---|---|---|
| | | Control | NoDR | DR |
| **True Label** | Control | 733 | 8 | 219 |
| | NoDR | 41 | 545 | 134 |
| | DR | 112 | 29 | 2109 |

In Table 2 and Table 3, the predictions of the CNN on Dataset 1 and Dataset 2, respectively, for three categories, Control, NoDR, and DR are shown. Based on these predictions, ACC, SE, and SP, are determined and summarized in Table 4 and Table 5.

*Table 4 Cross Validation Evaluation Metrics for OCTA dataset*

| n = 131 | Control | NoDR | DR | Average |
|---|---|---|---|---|
| **ACC (%)** | 87.022 ± 0.059 | 90.840 ± 0.020 | 83.970 ± 0.050 | 87.277 ± 0.034 |
| **SE (%)** | 78.123 ± 0.152 | 95.835 ± 0.089 | 77.334 ± 0.076 | 83.764 ± 0.105 |
| **SP (%)** | 89.899 ± 0.050 | 89.720 ± 0.022 | 92.858 ± 0.041 | 90.825 ± 0.018 |

*Table 5 Cross Validation Evaluation Metrics for augmented OCTA dataset*

| n = 3930 | Control | NoDR | DR | Average |
|---|---|---|---|---|
| **ACC (%)** | 90.331 ± 0.031 | 94.606 ± 0.039 | 87.430 ± 0.039 | 90.789 ± 0.036 |
| **SE (%)** | 76.354 ± 0.130 | 75.695 ± 0.174 | 93.733 ± 0.042 | 81.927 ± 0.102 |
| **SP (%)** | 94.849 ± 0.044 | 98.848 ± 0.010 | 78.988 ± 0.089 | 90.894 ± 0.105 |

Using the cross-validation predictions, ROC graphs were generated for both datasets in Fig. 3. Both graphs contain the ROC for the individual classes and the average for the model. The corresponding AUC was calculated for the individual classes and the average and is tabulated in Table 6.

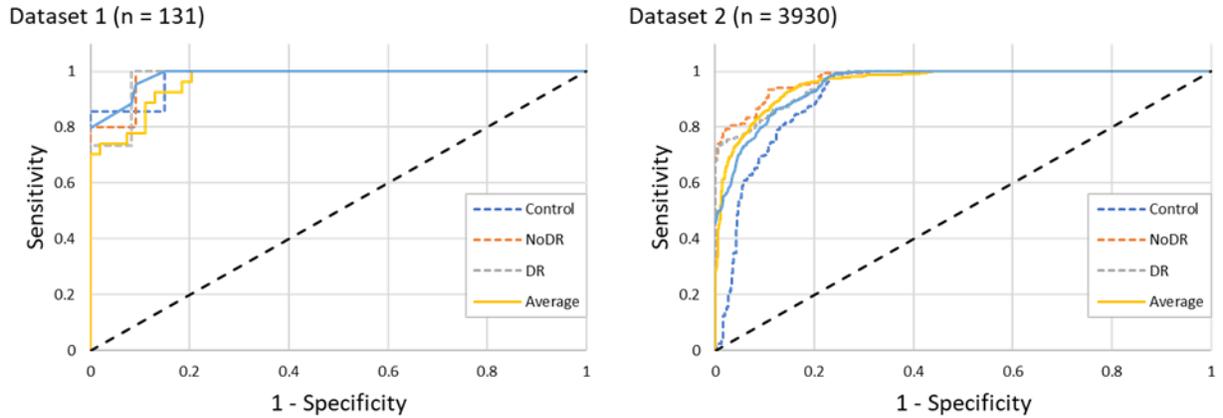

*Figure 3 ROC curves for the model's performance on (left) Dataset 1 (n = 131) and (right) Dataset 2 (n = 3930) for the individual class performance (control, NoDR, and DR) and the average performance of the model.*

*Table 6 AUC Performance*

| Dataset | Control | NoDR | DR | Average |
| --- | --- | --- | --- | --- |
| n = 131 | 0.97857 | 0.98182 | 0.97778 | 0.96502 |
| n = 3930 | 0.92072 | 0.97429 | 0.95715 | 0.95715 |

**Evaluation of Clinical Deployment**

The model with the best performance was integrated with a custom-designed GUI platform to evaluate the potential of the AI classification of DR using OCTA in a clinical environment. As shown in Fig. 4, The GUI adopted a commonly used feature used in retina clinics to enable easy adoption by ophthalmic personnel. By clicking the 'Load image', one OCTA image can be selected (Fig. 4A) and displayed (Fig. 4B) for visual examination. By clicking the 'Predict', automated AI classification can be executed. The output of the DR classification is displayed in the Results box. Additional information about the classification confidence is also available for clinical reference. The GUI platform has been tested by three ophthalmologists (Lim, Chan, and Toslak) to verify the feasibility of using the deep learning-based AI classification of DR using OCTA in a clinical environment. For each OCTA image, the GUI based classification can be completed within one minute.

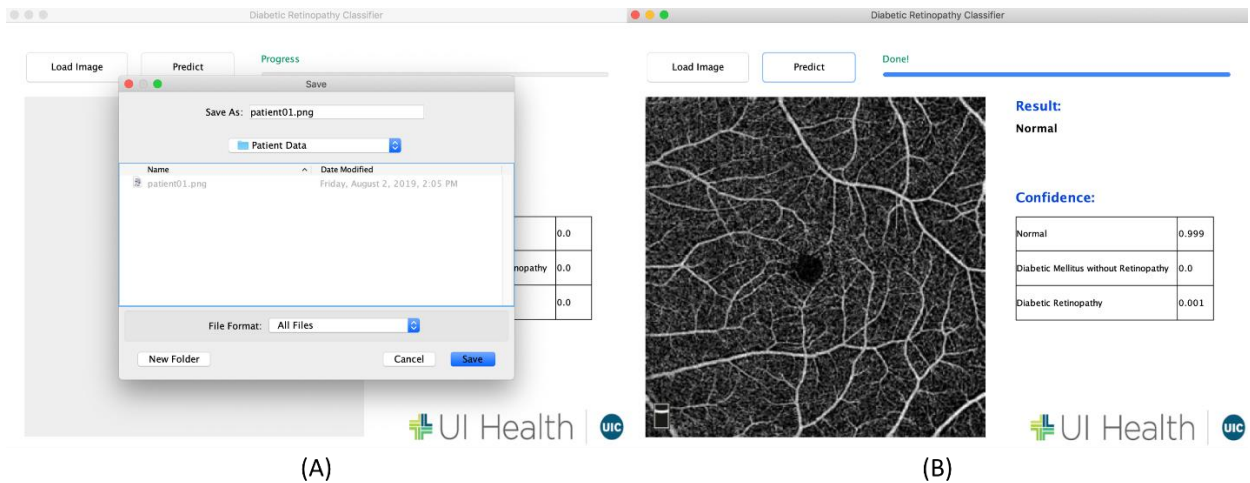

*Figure 4 GUI platform for DR classification using OCTA*

## DISCUSSION

In summary, we demonstrated the feasibility of using transfer learning for automated OCTA classification of DR. A model selection process was involved to evaluate and identify the number of upper layers required for re-training in the 16-layer CNN VGG16, to achieve optimal performance of OCTA classification. Moreover, a custom GUI platform was developed to verify the potential of using the AI classification in a clinical environment. The AI platform for automated OCTA classification may provide a practical solution to reduce the burden of experienced ophthalmologists for mass-screening of DR patients.

In principle, the CNN for deep learning can be adopted for OCTA classification of DR and other eye diseases. However, its practical application is challenging due to the limited datasets available for the relatively new OCTA imaging modality. The transfer learning process provides a solution to overcome this problem. By leveraging the weights of a pre-trained CNN, the OCTA data requirement can be reduced for clinical application. Since a CNN learns in a bottom-up hierarchy, we can keep using the bottom (early) layers and re-train the upper (later) layers to achieve rapid training for robust OCTA classification. In this study, we investigated the number of upper layers required for re-training in the 16-layer CNN VGG16 to achieve robust OCTA classification. For effective transfer learning, we measured the performance of OCTA classification as we incrementally increased the number of layers re-trained, starting with the final layer to the first (16 layers in total). Following the one standard deviation rule, we selected the simplest model that has non statistical differences, in terms of misclassification error, with the best performing model. Based on our study, for our OCTA dataset, the model re-trained with nine layers is the simplest model that achieves the lowest misclassification error. In Fig. 2, we compare the transfer learning performance of the model re-trained with nine layers and fifteen layers and illustrate that the misclassification error of the re-train nine layer is within one positive standard deviation of the re-train fifteen layer.

With the upper nine layers re-trained model, we conducted a cross-validation study to evaluate the model's performance. The cross-validation performance reveals that this model achieves an average of 87.277% ACC, 83.764% SE, and 90.825% SP across the three categories, control, NoDR, and DR when validating on Dataset 1. In comparison, for Dataset 2 the model achieves 90.789% ACC, 81.927% SE, and 90.894% SP. For the individual class predictions, in both datasets, NoDR had the highest accuracy, followed by Control, then DR. The performance is reflected in the ROC graphs and AUC. For Dataset 1, the overall AUC is 0.96, with individual AUC for control, NoDR, and DR as 0.97, 0.98, and 0.97 respectively. In comparison, the validation for Dataset 2, reveals more modest results with overall AUC as 0.95, and individual AUC for control, NoDR, and DR as 0.92, 0.97, and 0.95. Based on the results of the cross-validation study, robust performance of transfer learning for OCTA classification of DR was demonstrated. Using the selected model with nine upper layers re-trained, we exported the parameters of the model into a portable file and integrate it with a custom-designed GUI to assess the potential using the automated OCTA classification in a clinical environment.

There are some limitations for our current study. Transfer learning provided a reasonable performance for automated OCTA classification of DR. However, current OCTA datasets are limited. Furthermore, the dataset used in this study acquired from one device, and multi-device OCTA datasets are required for further validation of the transfer learning based OCTA classification.

22. Dave, V.P.; Pappuru, R.R.; Gindra, R.; Ananthakrishnan, A.; Modi, S.; Trivedi, M.; Harikumar, P. OCT angiography fractal analysis-based quantification of macular vascular density in BRVO eyes. *Canadian Journal of Ophthalmology* **2018**.

23. Alam, M.; Thapa, D.; Lim, J.I.; Cao, D.; Yao, X. Quantitative characteristics of sickle cell retinopathy in optical coherence tomography angiography. *Biomedical optics express* **2017**, *8*, 1741-1753.

24. Alam, M.; Zhang, Y.; Lim, J.I.; Chan, R.V.; Yang, M.; Yao, X. QUANTITATIVE OPTICAL COHERENCE TOMOGRAPHY ANGIOGRAPHY FEATURES FOR OBJECTIVE CLASSIFICATION AND STAGING OF DIABETIC RETINOPATHY *RETINA* **2019**.

25. Mo, S.; Krawitz, B.; Efstathiadis, E.; Geyman, L.; Weitz, R.; Chui, T.Y.; Carroll, J.; Dubra, A.; Rosen, R.B. Imaging foveal microvasculature: optical coherence tomography angiography versus adaptive optics scanning light ophthalmoscope fluorescein angiography. *Investigative ophthalmology & visual science* **2016**, *57*, OCT130-OCT140.

26. Alam, M.; Le, D.; Lim, J.I.; Chan, R.V.P.; Yao, X.C. Supervised Machine Learning Based Multi-Task Artificial Intelligence Classification of Retinopathies. *J Clin Med* **2019**, *8*, doi:ARTN 872 10.3390/jcm8060872.

27. Fang, L.; Cunefare, D.; Wang, C.; Guymer, R.H.; Li, S.; Farsiu, S. Automatic segmentation of nine retinal layer boundaries in OCT images of non-exudative AMD patients using deep learning and graph search. *Biomedical optics express* **2017**, *8*, 2732-2744.

28. Guo, Y.; Camino, A.; Wang, J.; Huang, D.; Hwang, T.S.; Jia, Y. MEDnet, a neural network for automated detection of avascular area in OCT angiography. *Biomedical optics express* **2018**, *9*, 5147-5158.

29. Kheradpisheh, S.R.; Ghodrati, M.; Ganjtabesh, M.; Masquelier, T. Deep networks can resemble human feed-forward vision in invariant object recognition. *Scientific reports* **2016**, *6*, 32672.
18